\documentclass[12pt]{article}
\usepackage{amsfonts,amsmath,epsfig,graphicx,amssymb}
\usepackage{amsmath,epsfig,graphicx,amssymb}
\usepackage[margin=1in]{geometry}
\usepackage{amsmath}
\usepackage{amsfonts}
\usepackage{amssymb}
\usepackage{makeidx}
\usepackage{graphicx}
\newcommand{\beq}{\begin{equation}}
\newcommand{\eeq}{\end{equation}}
\newcommand{\ben}{\begin{eqnarray}}
\newcommand{\een}{\end{eqnarray}}
\date{}
\begin{document}
\title{Context and Complementarity: Formalizing Bohr’s Vision through Many-Valued Contextual Logic}
\author{Partha Ghose \footnote{partha.ghose@gmail.com}\\Tagore Centre for Natural Sciences and Philosophy,\\ Rabindra Tirtha, New Town, Kolkata 700156, India\\ and\\Sudip Patra\\
Center for Complexity Economics, Applied Spirituality\\ and Public Policy (CEASP), O P Jindal Global University,\\ Sonipat, Haryana 131001, India} 
\maketitle
\begin{abstract}
Quantum mechanics challenges classical intuitions of space, time, and causality through the superposition principle, which allows systems to exist in multiple states simultaneously. Niels Bohr responded to these paradoxes with his {\em Complementarity Principle}: certain properties--such as wave and particle behaviours--are mutually exclusive yet jointly necessary for a full description of quantum systems. Crucially, Bohr insisted that the measuring apparatus must be regarded as an intrinsic part of any phenomenon, introducing two foundational ideas:

{\em Complementarity}: Different experimental setups reveal different, mutually exclusive aspects of reality.

{\em Contextuality}: Measurement outcomes depend on the broader experimental context.

Experiments such as the double-slit setup and theoretical results like the Kochen-Specker theorem reinforce the view that quantum observables lack pre-existing values independent of measurement. Bohr, influenced by thinkers such as Edgar Rubin and William James, also drew analogies with psychology, emphasizing the inseparability of observer and observed.

However, Bohr did not provide a systematic logical framework for complementarity and contextuality. To address this, Hans Reichenbach proposed a three-valued logic that was {\em implicitly} contextual, while da Costa and Krause developed paraconsistent logics to manage contradictions without collapse into triviality.

In this paper, we present a generalization of Reichenbach's logic--a {\em seven-valued, formally contextual system} in which contextuality is explicitly captured using existential quantifiers. The indeterminate third value remains philosophically akin to Reichenbach's original conception, while additional values model conditional truth under different, incompatible measurement arrangements. This approach accommodates quantum paradoxes--such as the double-slit, Schr\"{o}dinger's cat, Wigner's friend, and the EPR argument--without sacrificing logical coherence or requiring nonlocal explanations.

Our framework draws philosophical inspiration from Jain logic, particularly {{\em {sy\={a}dv\={a}da}}}, which endorses multiple, context-dependent truths. This pluralistic logic not only deepens our understanding of quantum theory but also resonates with challenges in fields like perceptual psychology and cognitive science, where context-dependence and apparent contradictions similarly abound.

This work seeks to extend Bohr's philosophical legacy by providing the formal logical tools that his foundational insights demand.

\end{abstract}

\section{Introduction}
Quantum mechanics is filled with paradoxes that challenge our usual ways of understanding the world in terms of space, time, and causality. They originate in the quantum superposition principle. Niels Bohr addressed this puzzling nature of quantum theory through his Complementarity Principle, a key idea introduced in the late 1920s. According to this principle, certain physical properties—like position and momentum, or wave and particle behaviour—are mutually exclusive: they cannot be observed at the same time. Yet both are essential to fully describe quantum systems. The catch is that each of these properties can only be revealed in specific, incompatible experimental setups.
As Bohr put it \cite{bohr}:

``The very nature of the quantum theory forces us to regard the measuring apparatus as an intrinsic part of any phenomenon we describe.''

This quote captures Bohr's central insight: in quantum mechanics, what we observe depends on how we observe it. The measuring device is not just a passive tool—it actively shapes the outcome. This leads directly to two key concepts:

•	Complementarity: different experiments reveal different, mutually exclusive aspects of quantum systems.

•	Contextuality: the outcome of a measurement depends on the broader experimental context.
 
For instance, in the double-slit experiment, an electron behaves like a wave when we do not observe which path it takes, but like a particle when we do. Each setup provides only part of the full picture.

The idea of contextuality has been formalized in the Kochen-Specker theorem \cite{ks}, which shows that it is impossible to assign fixed, non-contextual values to all quantum properties. In short, quantum mechanics does not describe a world where observables have definite values independent of how we measure them. This aligns with Bohr’s philosophical view: the act of measurement is inseparable from the phenomenon observed.

Importantly, Bohr also saw parallels between quantum theory and fields beyond physics, such as psychology. He was influenced by thinkers like Edgar Rubin and William James, who emphasized the interdependence of subject and object. Bohr's reflections on the parallels between quantum physics and psychology may be paraphrased as: ``In psychology, we are confronted with a situation where the subject and the object cannot be sharply separated.'' This is analogous to quantum physics, where the observer and the observed system are entangled. Just as introspection alters the mental state being observed in psychology, any observation in quantum mechanics influences the system being measured.

However, Bohr never formalized his idea of complementarity within a formal logical framework, which has led to ongoing debates and differing interpretations and the exploration of non-classical logics. The first formal departure from classical logic in response to quantum mechanics came with Birkhoff and von Neumann’s `quantum logic' \cite{bir}, which proposed a lattice-theoretic structure to capture the non-classical propositional relationships observed in quantum theory. This logic replaces the Boolean algebra of classical propositions with an orthomodular lattice, reflecting the failure of the distributive law in quantum contexts. While foundational, their approach was largely syntactic and algebraic, without addressing contextuality or observer dependence in an explicit epistemological sense. It thus differs markedly from later contextual and paraconsistent logics, such as those by Reichenbach \cite{reich} or da Costa \cite{costa}, which aim to integrate logical frameworks more directly with the conceptual demands of quantum measurement and epistemology. The works of Reichenberg, Gardner \cite{gard}, da Costa and de Ronde \cite{ronde}, Becker Arenhart and Krause \cite{bec}, and Estrada-Gonz\'alez and Cano-Jorge \cite{est} offer distinct yet interrelated perspectives on this subject.

Hans Reichenberg's seminal work ``Philosophic Foundations of Quantum Mechanics'' introduces a three-valued logical system to interpret quantum mechanics. His logic incorporates an indeterminate truth value to account for the probabilistic nature of quantum events, providing an alternative to classical binary logic and aiming to reconcile the apparent inconsistencies in quantum observations. 

Gardner's ``Two Deviant Logics for Quantum Theory: Bohr and Reichenberg'' offers a comparative analysis of Bohr's complementarity principle and Reichenberg's three-valued logic. Gardner elucidates how both frameworks deviate from classical logic to resolve quantum paradoxes, highlighting their respective approaches to addressing the challenge posed by quantum phenomena.

In `The Paraconsistent Approach to Quantum Superpositions Reloaded: Formalizing Contradictiory Powers in the Potential Realm', da Costa and de Ronde proposed a paraconsistent approach to quantum superpositions. As they note, ``Bohr attempted to avoid contradictions through the epistemological notion of complementarity, which restricts the applicability of classical concepts to experimental arrangements. However, we explore the possibility that contradictions are real and can be formally accommodated via paraconsistent logic.'' In contrast to Bohr's epistemic stance, their approach rejects the necessity of avoiding contradictions and does not appeal to contextuality as a means of resolving them. Instead, it embraces contradictions as genuine features of quantum systems, to be managed through formal logical frameworks. 

In ``Potentiality and Contradiction in Quantum Mechanics'' Becker Arenhart and Krause critically examine the proposition that  quantum supepositions embody real contradictions. They argue against the notion that superposed stats represent contradictory properties, suggesting instead that such states should be viewed as `contraries' rather than `contradictions'. This perspective challenges interpretations that employ paraconsistent logic to accommodate contradictions within quantum mechanics.

In ``A Defense of the Paraconsistent Approach to Quantum Superpositions (Answer to Arenhart and Krause)' Christian de Ronde defends the `Paraconsistent Approach to Quantum Superpositions' (PAQS) in response to criticisms by Arenhart and Krause. He challenges the assumption that the interpretation of quantum mechanics must be governed by classical logic, emphasizing instead a crucial distinction between the actual realm—-where classical properties are instantiated—-and the potential realm, where quantum superpositions reside. Within this potential realm, de Ronde argues, contradictions can meaningfully exist without undermining consistency at the level of actuality. On this basis, PAQS offers a coherent metaphysical framework that aligns with the formalism of quantum mechanics and allows for a deeper understanding of quantum phenomena without reducing them to classical terms. De Ronde thus calls for a shift away from classical metaphysical constraints and toward a new ontology that accommodates the non-classical, potentially contradictory character of quantum reality.

Estrada-Gonz\'alez and Cano-Jorge, in ``Revisiting Reichenbach's Logic'', analyze Reichenbach's three-valued logic using contemporary logical tools. They demonstrate that Reichebach's approach, when viewed through Dunn-style semantics, reveals features that address previous criticisms and establish connections with other non-classical logics. Their work underscores the relevance of Reichenbach's logic in modern discussions on the logical foundations of quantum mechanics.

Collectively, these works contribute to the discourse on the applicability of non-classical logics in quantum mechanics. While Reichenbach's three-valued logic and Bohr's complementarity represent early efforts to address quantum paradoxes, contemporary analyses by da Costa and de Ronde, Becker Arenhart, Krause, Estrada-González, and Cano-Jorge offer critical evaluations and modern reinterpretations of these foundational ideas. Their contributions enrich the ongoing dialogue on how best to conceptualize and formalize the counterintuitive aspects of quantum theory.

Our work can be seen as a further contribution to these lines of thought. It proposes a non-classical seven-valued logic that is formally contextual. While the third truth-value (indeterminate) bears resemblance to Reichenbach's, our approach is a generalization of his three-valued logic to a seven-valued system in which contradictions are resolved by explicitly {\em formalizing} contextuality. In doing so, it remains consistent with Bohr's insights and extends them within a coherent and formally articulated logical framework.

Furthermore, unlike earlier approaches, our framework is not limited to resolving quantum paradoxes but is designed to be applicable across a range of domains beyond quantum physics. It draws on Jain logic—-particularly {\em sy\={a}dv\={a}da}, which embraces multiple, context-dependent truths--to offer a pluralistic philosophical system. In other words, our aim is to bridge quantum theory with other complex fields, such as psychology and sociology, enabling a broader and more integrated understanding of diverse phenomena.

\section{A Many-valued Contextual Logic}

Many-valued logic systems generalize the familiar two-valued (true/false) framework of classical logic. Developed primarily in the 20th century by European and American logicians such as Jan Lukasiewicz, C. I. Lewis, Stephen Kleene, Emil Post, and Bochvar, these systems tend to be highly technical and abstract in nature. One notable exception is fuzzy logic--an infinite-valued logic with continuous gradations of truth--which has found widespread practical application in fields like engineering, pattern recognition, and medicine.

In contrast to these formal, mathematically grounded systems, many-valued logic also has a rich philosophical lineage outside the Western tradition. An illuminating example is the ancient Jaina logical system known as {{\em {sy\={a}dv\={a}da}}}. Unlike Western formal logics, which typically seek universal truth-values abstracted from context, {\em {sy\={a}dv\={a}da}} is explicitly and intrinsically contextual. The key term {\em {sy\={a}t}} has been variously translated as `perhaps', `may be', `in some way', or `under certain conditions'. From the perspective of physics and experimental psychology, the most fitting interpretation is `in a certain context'.

A classic illustration used by Jaina thinkers involves the transformation of a clay pot: before firing, the pot is black; after firing, it is red; and during the firing process, its colour is {\em avaktavyam}--a term denoting what is unspeakable, indeterminate, or undecidable. In this way, {\em {sy\={a}dv\={a}da}} introduces a third fundamental truth value beyond true (T) and false (F), namely {\em avaktavyam} (indeterminate) (U). These are always conditionally asserted, prefixed by {\em {sy\={a}t}} to indicate contextuality:

{\em {sy\={a}t asti}} – it is (true, in some respect),

{\em {sy\={a}t n\={a}sti}}  – it is not (false, in some respect),

 {\em {sy\={a}t avaktavyam}} – it is indeterminate (in some respect).

These three core values generate a richer system known as {\em saptabhaṅgīnaya}, or the sevenfold schema of predication, encompassing all possible logical combinations of T, F, and U, each explicitly contextualized:

{\em sy\={a}t  asti cha n\={a}sti cha}  – it is and it is not (true and false, in some respect),

{\em sy\={a}t asti cha avaktavyam cha} – it is and it is indeterminate (true and indeterminate, in some respect),

{\em sy\={a}t n\={a}sti cha avaktavyam cha}  – it is not and it is indeterminate (false and indeterminate, in some respect),

{\em sy\={a}d asti cha n\={a}sti cha avaktavyam cha}  – it is, it is not, and it is indeterminate (true, false, and indeterminate, in some respect). The Sanskrit word {\em cha} means `and'. 

Unlike modern many-valued logics, which operate within a fixed logical space, {\em {sy\={a}dv\={a}da}} underscores the conditional and perspectival nature of all assertions. It is not merely a system of truth-values, but a philosophical orientation toward reality's complexity, ambiguity, and dependence on standpoint. In that sense, it represents a fundamentally different--yet profoundly sophisticated--approach to logic.

Drawing inspiration from this system, a many-valued logic has been proposed which is formally contextual \cite{ghose}. Using the quantifier $\forall$ the three basic conditional truth values can be written as

(i) $\forall x\, [\phi(x) \rightarrow p(x)]$; 

(ii) $\forall x\, [\phi(x) \rightarrow \neg p(x)]$; 

(iii) $\forall x\, [\phi(x) \rightarrow q(x)]$.

Here, $x$ is a variable ranging over a domain of discourse (e.g., clay pots), $\phi$ is a well-formed formula specifying a context or condition (such as `is baked'), $p$ is a predicate (e.g., `is red'), and $q$ is the predicate expressing indeterminacy.

To illustrate: the first formula, $\forall x\, [\phi(x) \rightarrow p(x)]$ can be read in plain English as, `For all $x$ (say, clay pots), if $x$ satisfies the condition $\phi$ (e.g., is baked), then $x$ is red'. The logic retains explicit reference to the condition under which the truth of a statement is evaluated, capturing the essence of contextual truth central to {\em {sy\={a}dv\={a}da}}.

The other four compounds may be written as 

(iv) $\forall x\,[\phi(x) \rightarrow p(x) \land \phi^\prime(x)\rightarrow \neg p(x)] \land \neg [ \phi(x) \leftrightarrow \phi^\prime (x)]$,  

(v) $\forall x\,[\phi(x) \rightarrow p(x) \land \phi^\prime(x)\rightarrow q(x)] \land \neg [ \phi(x) \leftrightarrow \phi^\prime (x)]$,  

(vi) $\forall x\,[\phi(x) \rightarrow \neg p(x) \land \phi^\prime(x)\rightarrow q(x)] \land \neg [ \phi(x) \leftrightarrow \phi^\prime (x)]$,  
 
(vii) $\forall x\,[\phi(x) \rightarrow p(x) \land \phi^\prime(x)\rightarrow \neg p(x) \land \phi^{\prime\prime}(x) \rightarrow q(x)] \land \neg [\phi(x) \leftrightarrow \phi^\prime (x)] \land\neg[ \phi^\prime (x) \leftrightarrow \phi^{\prime\prime}(x)] \land \neg[\phi(x)\leftrightarrow \phi^{\prime\prime}(x)]$.

When expressed in this formal framework, the seven predications--comprising the combinations of three basic truth-values—-are mutually consistent, as each holds under distinct, non-overlapping conditions.

A more compact notation for these would be:

(i) $c T$,

(ii) $c F$, 

(iii) $c U$, 

(iv) $c_1 T$ and $c_2 F$, 

(v) $c_1 T$ and $c_2 U$, (vi) $c_1 F$ and $c_2 U$, 

(vii) $c_1 T$ and $c_2 F$ and $c_3 U$, 

where each $c$ represents a specific condition or context (a $\phi(x)$) under which the truth-value is asserted.

In classical logic, contradictions lead to the principle of explosion--{\em ex contradictione quodlibet}--meaning that from a contradiction, any proposition can be inferred. Formally contextual logics as well as paraconsistent logics, however, are specifically designed to block this explosion. They allow contradictions to exist without collapsing the entire logical system, precisely because they are propositionally weaker and more restrained than classical logic.

Graham Priest has shown \cite{graham} that {\em {sy\={a}dv\={a}da}} can, in fact, also be rigorously reconstructed within a paraconsistent logical framework.

Paraconsistent logics--such as those developed by da Costa, Belnap \cite{bel}, and Priest--focus on preserving consistency in the face of contradictions by weakening certain inference rules, especially to avoid the principle of explosion. These systems are primarily syntactic or semantic innovations within the broader tradition of formal Western logic, applied to domains like inconsistent mathematics, legal reasoning, and database theory.

Our approach is distinct in both inspiration and formulation. It is philosophically rooted in the idea of {\em contextual and perspectival truth}. However, instead of incorporating the qualifier {\em {sy\={a}t}} directly, we formalize contextuality using existential and universal quantifiers over conditions--i.e., statements of the form $\forall x\, [\phi(x) \rightarrow p(x)]$--where the condition $\phi(x)$ plays the role of contextual grounding.

This marks a significant departure from {\em {sy\={a}dv\={a}da}}:: whereas {\em {sy\={a}dv\={a}da}} relies on the linguistic and epistemological qualifier {\em {sy\={a}t}} to assert propositions conditionally, our logic encodes context through a quantificational framework. This yields a formally precise system that captures the spirit of {\em {sy\={a}dv\={a}da}}--especially its embrace of indeterminacy and contradiction--while diverging from its original linguistic form. In this way, our approach complements classical and Jaina logics by formalizing context-sensitivity using the apparatus of modern logic.

\section{Many-valued Contextual Logic and Quantum Mechanics}

{{\flushleft{\em The Double-slit Paradox}}}

Contextuality plays a fundamental role in quantum mechanics. The famous double-slit experiment exemplifies this in its essentials--with both slits in an opaque plate open, electrons from an electron gun (or photons, neutrons or any atomic objects) passing through the apparatus one at a time produce an interference pattern on a distant screen over a period of time (wave-like behaviour), but with only one slit open or if the electrons are detected near one of the slits, this interference pattern disappears and a particle-like behaviour is observed. This is commonly known as `wave-particle complementarity' \cite{bohr}. The two experimental set ups here are {\em mutually exclusive}, and they give rise to {\em mutually incompatible} observations in terms of classical concepts like waves and particles. They can, however, be interpreted as {\em complementary aspects} of the same physical reality revealed in {\em mutually exclusive} `experimental conditions', or in other words, mutually exclusive `measurement contexts' which include `preparation' and `observation of the system'. 
\begin{figure}[h]
\centering
{\includegraphics[scale=0.6]{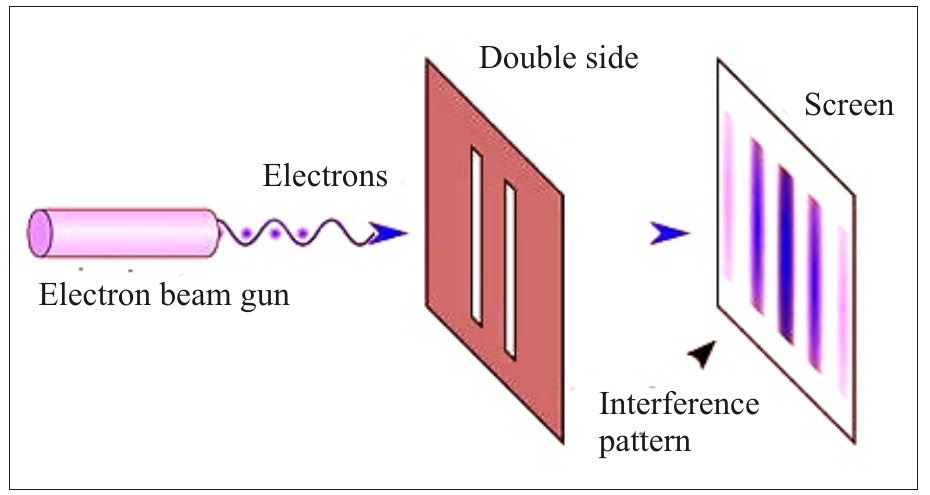}}
\caption{\label{Figure 1}{\footnotesize Schematic diagram of a double-slit experiment. electrons from an electron gun pass through a plate with two narrow slits, one at a time, and produce an interference pattern on a distant screen over a period of time, showing wave-like behaviour. If one of the slits is closed, or the electron is detected near one slit, this pattern disappears and a single-slit pattern appears, showing particle-like behaviour.}}
\end{figure}
We will adopt this as the basic definition of contextuality in quantum mechanics.

Let us take the double-slit experiment as an exemplar and see how the seven predications provide a complete list of logically independent variations of the double-slit set up:

(i) $\forall x\, [\phi(x) \rightarrow p(x)]$ corresponds to ``for all quantum mechanical entities like electrons, photons, atoms, etc, if $\phi(x)$ holds, i.e. say if only one slit is open and the entity is observed/detected, it is a `particle' (i.e. p(x) is the case).

(ii) $\forall x\, [\phi (x) \rightarrow \neg p(x)]$ corresponds to ``for all quantum mechanical entities, if $\phi (x)$ holds, i.e. say if only one slit is open and the entity is not observed/detected, the entity is not a `particle' (i.e. $\neg p(x)$ is the case). In fact, this produces a single-slit diffraction pattern characteristic of a wave.

(iii) $\forall x\, [\phi^\prime(x) \rightarrow q(x)]$ corresponds to ``for all quantum mechanical entities, if $\phi^\prime(x)$ holds, i.e. say if both slits are open and no observation is made, the entity is in a pure state (i.e. $q(x)$ is the case). 

(iv) $\forall x\, [\phi(x) \rightarrow p(x) \land \phi^\prime(x)\rightarrow \neg p(x)] \land \neg [ \phi(x) \leftrightarrow \phi^\prime (x)]$ corresponds to ``for all quantum mechanical entities, if $\phi(x)$ holds (say only one slit is open and the entity is observed/detected), it is a particle ($p(x)$ is the case), and under a mutually incompatible condition $\phi^\prime(x)$ (say both slits are open and no observation is made) the entity is not a particle ($\neg p(x)$ is the case).

(v) $\forall x\,[\phi(x) \rightarrow p(x) \land \phi^\prime(x)\rightarrow q(x)] \land \neg [\phi(x) \leftrightarrow \phi^\prime (x)]$ corresponds to ``all quantum mechanical entities are particles when $\phi(x)$ holds (say a single slit is open and the entity is observed/detected), and under a mutually incompatible condition $\phi^\prime(x)$ (say both the slits are open and no observation is made) they are in a pure state ($q(x)$ is the case). 

(vi) $\forall x\,[\phi (x) \rightarrow \neg p(x) \land \phi^\prime(x)\rightarrow q(x)]\land \neg [ \phi(x) \leftrightarrow \phi^\prime(x)]$,  
 corresponds to ``for all quantum mechanical entities if $\phi (x)$ holds (say a single slit is open and no observation is made), the entity is not a particle ($\neg p(x)$ is the case), and if the mutually incompatible condition $\phi^\prime (x)$ holds (say both slits are open and no observation is made), they are found in a pure state ($q(x)$ is the case). This corresponds to single-slit diffraction and double-slit interference patterns.

(vii) $\forall x\, [\phi(x) \rightarrow p(x) \land \phi^\prime(x)\rightarrow \neg p(x) \land \phi^{\prime\prime}(x) \rightarrow q(x)] \land \neg [\phi(x) \leftrightarrow \phi^\prime (x)] \land\neg[ \phi^\prime (x) \leftrightarrow \phi^{\prime\prime}(x)] \land \neg[\phi(x)\leftrightarrow \phi^{\prime\prime}(x)]$ corresponds to ``all quantum mechanical entities are particles under one type of condition $\phi(x)$ (a single slit is open and an observation is made), not particles under a mutually incompatible condition $\phi^\prime (x)$ (a single slit is open but no observation is made), and also are in pure states ($q(x)$ is the case) under yet another condition $\phi^{\prime\prime} (x)$ (both slits are open and no observation is made) which is mutually incompatible with the first two conditions. 
With three basic truth values, the number of possible combinations is $2^3 -1 = 7$, and hence these seven combinations (predications) exhaust all logically independent possibilities.
  
The seven logically independent predications collectively offer the most comprehensive description of a quantum mechanical entity and furnish the logical framework underlying Bohr's {\em Principle of Complementarity}.

Bohr formulated the principle in the following general way \cite{bohr}:

`Information regarding the behaviour of an atomic object obtained under definite experimental conditions may, however, according to a terminology often used in atomic physics, be adequately characterized as {\em complementary} to any information about the same object obtained by some other experimental arrangement excluding the fulfillment of the first conditions. Although such kinds of information {\em cannot be combined into a single picture} by means of ordinary concepts, they represent indeed equally essential aspects of any knowledge of the object in question which can be obtained in this domain'. (second italics inserted) 

{{\flushleft{\em The Schr\"{o}dinger Cat Paradox}}}

\begin{figure}[h]
\centering
{\includegraphics[scale=0.7]{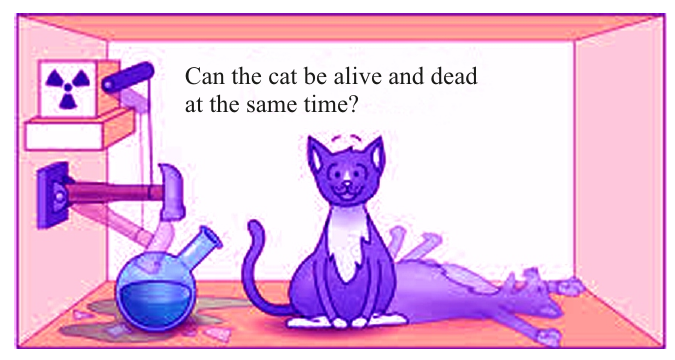}}
\caption{\label{Figure 2}{\footnotesize Cartoon of the Schr\"{o}dinger cat in a sealed box containing a radioactive substance, a Geiger counter, a hammer and a flask containing a poisonous gas. If the radioactive substance emits a radiation (an entirely random event) within a stipulated time, the Geiger counter will detect it, break the flask and release the gas which will kill the cat. If not, the cat will be alive. Until the box is opened, according to quantum mechanics, the cat will be in a superposition of live and dead states.}}
\end{figure}

This is a famous paradox in quantum mechanics. It is about a hypothetical cat (representing a macroscopic quantum mechanical object) locked up in a box with a radioactive substance inside, a Geiger counter to detect any radiation emitted by the substance, a hammer and a flask containing a poisonous gas which will kill the cat if released \cite{schr}. Radioactive decays are purely random events.
Let the probability that the radioactive substance emits a radiation within one hour of locking up the cat be 1/2. If the radiation is emitted within this time, the Geiger counter will detect the radiation and cause the hammer to break the flask and release the poisonous gas which will kill the cat. According to quantum mechanics the cat will be in a superposition of live and dead states after an hour:
\beq
|Cat\rangle = \frac{1}{\sqrt{2}}[|Cat\rangle_{alive} + |Cat\rangle_{dead}]
\eeq
But if one opens the box after an hour, the cat will be found to be either dead or alive with $50 \%$ probability. How can a cat be neither dead nor alive when not observed? And, how can it collapse to one of these states when observed? This is the paradox. Schr\"{o}dinger named such a state a `grotesque' state.

The resolution lies in recognizing it to be instances of predications (v) or (vi) depending on whether the cat is found alive or dead on opening the box. Predication (v) is $c_1 T$ and $c_2 U$ ($\forall x\,[\phi(x) \rightarrow p(x) \land \phi^\prime(x)\rightarrow q(x)] \land \neg [ \phi(x) \leftrightarrow \phi^\prime (x)]$). The simultaneously alive and dead cat is in a superposition of states $q(x)$ when inside the sealed box (condition $\phi^\prime (x)$ applies) and the cat is alive ($p(x)$) when the box is opened and someone looks inside (condition $\phi(x)$ applies). There is no contradiction because the conditions $\phi(x)$ and $\phi^\prime (x)$ are mutually exclusive and incompatible. 

Predication (vi) is $c_1 F$ and $c_2 U$ ($\forall x\,[\phi(x) \rightarrow \neg p(x) \land \phi^\prime(x)\rightarrow q(x)] \land \neg [ \phi(x) \leftrightarrow \phi^\prime (x)]$). It corresponds to the cat being in a superposition of states ($q(x)$) when inside the box (condition $\phi^\prime (x)$ applies) and dead ($\neg p(x)$) when the box is opened and someone looks inside (condition $\phi(x)$ applies). Again, there is no contradiction because $\phi(x)$ and $\phi^\prime (x)$ are mutually exclusive and incompatible conditions. 
  
If one admits that there is no objectively real and noncontextual state of a quantum system, there is no paradox.

{{\flushleft{\em Wigner's Friend Paradox}}}

Another vexing quantum paradox is that of Wigner's friend \cite{wig}. Eugene Wigner thought of this paradox to bring to sharp focus the tension between an objective quantum state and its objective collapse on measurement as envisioned by the Copenhagen school (barring Bohr). Let $F$ be Wigner's friend who is inside a laboratory performing an experiment with a quantum system $S$ (say a spin 1/2 particle). According to standard quantum mechanics, both Wigner and his friend $F$ are also quantum mechanical systems. Before any measurement, $S$ is in a superposition of states
\beq
|S\rangle = \frac{1}{\sqrt{2}}\left[|\rm{spin\, up}\rangle + |\rm{spin\, down}\rangle \right]
\eeq
Suppose after the measurement $F$ finds the `spin up' state. Then, applying the collapse postulate, she will write the state after measurement as
\beq
|S^\prime\rangle = |\rm{spin\, up}\rangle
\eeq
The state $|S\rangle$ has collapsed to $|S^\prime\rangle$.

If Wigner ($W$) is outside the lab and is informed that his friend $F$ has made a measurement but is not told which result she  has obtained, he will write down the state
\beq
|S\rangle \frac{1}{\sqrt{2}} \left[|F(\rm{up})\rangle +  |F(\rm{down})\rangle\right]
\eeq
where $F(\rm{up})$ and $F(\rm{down})$ are the two possible states of F, i.e. F finding `spin up' or `spin down', each with $50\%$ probability.
To $W$ the original state $S$ has not collapsed! This is the paradox. This has led to many interpretations and controversies \cite{west}. It is clear that the paradox disappears if one regards all states as contextual rather than objectively real and noncontextual. The relevant predication is (v) $c_1 T$ and $c_2 U$ ($\forall x\,[\phi(x) \rightarrow p(x) \land \phi^\prime(x)\rightarrow q(x)] \land \neg [\phi(x) \leftrightarrow \phi^\prime (x)]$). For $F$ the context is $\phi(x)$ and the result is $p(x)$ (`spin up'), whereas for $W$ the context is $\phi^\prime(x)$ and the result is $q(x)$, a pure or {\em avaktavyam} state, and these two contexts are mutually exclusive and incompatible.  

{{\flushleft{\em The EPR Paradox}}}
 
Contextuality also plays a fundamental role in resolving the famous EPR paradox \cite{epr}. Instead of following the rather long and involved arguments given in Ref. \cite{epr}, let us follow the much more succint and clear exposition given by Einstein later \cite{eins}. Let two systems A and B interact and go far apart so that there is no significant interaction between them any more. To paraphrase Einstein, let each system be a qubit, i.e. a superposition of two orthogonal basis states $(A_0, A_1)$ and ($B_0, B_1$). The qubits can alternatively be written as superpositions of two other basis states ($A_+, A_-$) and ($B_+, B_-$) which are linear combinations of $(A_0, A_1)$ and ($B_0, B_1$) respectively. Quantum mechanically the two representations are equivalent. Quantum mechanics predicts that when they are far apart, they will form an entangled (i.e. nonfactorizable) state like
\ben
\Psi &=& \frac{1}{\sqrt{2}}[A_0 B_0 + A_1 B_1]\label{ent1}\\
&=& \frac{1}{\sqrt{2}}[A_+ B_+ + A_- B_-]\label{ent2}
\een
in which the subsystems A and B no longer have a definite state of their own although the total system does have a well defined state. Now, if Alice makes a measurement on A in the (0,1) basis, she will find A in the state $A_0$ with $50\%$ probability, and this will result in B being in the state $B_0$ in accordance with the state (\ref{ent1}). On the other hand, if Alice were to make a measurement on A in the (+,-) basis, she would find it in the state $A_+$ with $50\%$ probability, and correspondingly B will be in the state $B_+$ in accordance with the state (\ref{ent2}). Thus, mutually exclusive measurement choices on the system A result in two different states for the system B. Einstein argued that `on one supposition we should, in my opinion, absolutely hold fast': the `real factual situation' of the system B (which he calls $S_2$) must be independent of what is done with A (which he calls $S_1$) which is spatially separated from A. Einstein then argues that for the same real situation of B it is possible therefore to find, according to one's choice, different types of wave functions or states for B. He goes on to write, `One can escape from this conclusion only by either assuming that the measurement of $S_1$ ( (telepathically) ) changes the real situation of $S_2$ or by denying independent real situations as such to things which are spatially separated from each other. Both alternatives appear to me entirely unacceptable'. Hence he concludes that, since two different states or wave functions can be coordinated with the identical real factual situation of B, quantum mechanics cannot be a complete description of nature.

Insisting on a `real factual situation' of a system irrespective of any measurement context is tantamount to assuming a reality that is noncontextual. Bohr pointed out that the problem (or paradox) is resolved if one views reality as contextual, i.e. dependent on the measurement context \cite{bohr2}. Elsewhere Bohr expressed the impossibility of separating the behaviour of a quantum system from the measuring apparatus in the following words \cite{bohr3}:

``This crucial point ... implies the impossibility of any sharp separation between the behaviour of atomic objects and the interaction with the measuring instruments which serve to define the conditions under which the phenomena appear. In fact, the individuality of the typical quantum effects finds its proper expression in the circumstance that any attempt of subdividing the phenomena will demand a change in the experimental arrangement introducing new possibilities of interaction between objects and measuring instruments which in principle cannot be controlled. Consequently, evidence obtained
under different experimental conditions cannot be comprehended within a single picture, but must be regarded as complementary in the sense that only the totality of the phenomena exhausts the possible information about the objects.'' 
 
This is Bohr's `Complementarity Principle'-- the incompatible results in the two mutually exclusive measurement contexts provide complementary descriptions of the system which together provide a more comprehensive view of reality. Thus, {\em incompleteness of the theory based on noncontextual reality turns into a more comprehensive completeness based on contextual reality}. We have already seen in detail how the seven-valued contextual logic developed above collectively furnishes the logical foundation of Bohr's 'Complementarity Principle'.

Contextuality can also be defined in terms of hidden variable theories \cite{ks}, but we will not discuss that here. 
 
{{\flushleft{\em Quantum Cheshire Cat}}}

There is a rather intriguing paradox known as the Quantum Cheshire Cat (QCC) paradox. In 2013 Aharanov and his colleagues proposed a gedanken experiment in which they found `the Cat in one place, and its grin in another. The Cat is a photon, while the grin is its circular polarization' \cite{aha}. The allusion is to Lewis Carroll's {\em Alice's Adventures in Wonderland} where one finds:

{\em `All right,' said the Cat; and this time it vanished quite slowly, beginning with the end of the tail, and ending
with the grin, which remained some time after the rest of it had gone.

`Well! I've often seen a cat without a grin,' thought Alice, `but a grin without a cat! It's the most curious thing I
ever saw in my life!'}

Aharanov {{\em et al} go on to write:

`In real life, assuming that cats do indeed grin, the grin is a property of the cat; it makes no sense
to think of a grin without a cat. And this goes for almost all physical properties. Polarization is a property of photons; it makes
no sense to have polarization without a photon. Yet, as we will show here, in the curious way of quantum mechanics, photon
polarization may exist where there is no photon at all. At least this is the story that quantum mechanics tells via measurements
on a pre- and post-selected ensemble.'

\begin{figure}[h]
\centering
{\includegraphics[scale=0.7]{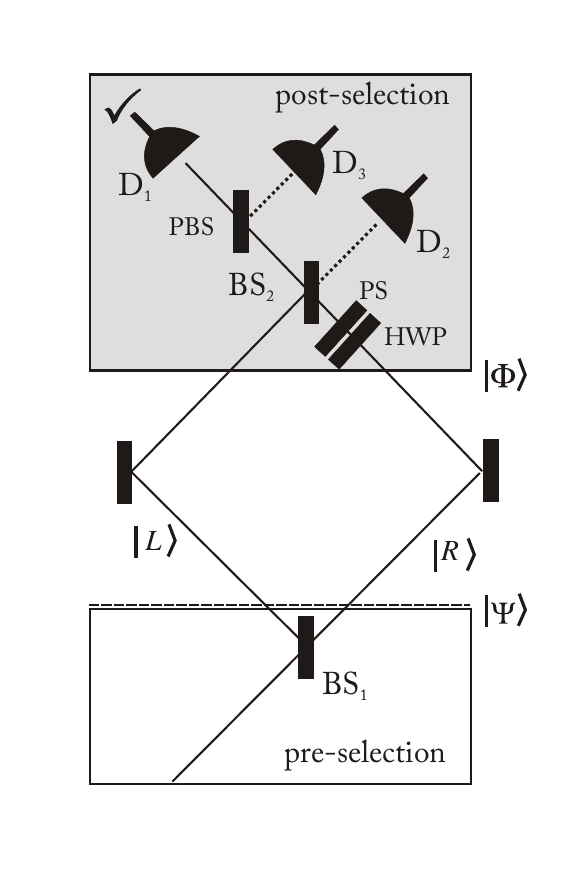}}
\caption{\label{Figure 3}{\footnotesize Schematic diagram of setup. A $H$ photon travels two paths $L$ and $R$ after the first beam splitter $BS_1$. This is the `pre-selected' state $|\Psi\rangle$. The photons along the two arms are then combined on a second beam splitter $BS_2$ after which two ideal detectors $D_1$ and $D_2$ are placed along the direction of the $R$ and $L$ arms respectively. $D_2$ is the bright port and $D_1$ is the dark port. If now a half-wave plate $HWP$ is placed in the path $R$ just before $BS_2$ followed by a phase shifter $PS$ which adds a phase $\pi/2$, it flips the polarization from $H$ to $V$ and produces the `post-selected' state $|\Phi\rangle$. Various other measuring devices are inserted into the set up, between the pre- and post-selection, as in Ref. \cite{aha}}}
\end{figure} 
Let us consider a Mach-Zehnder interferometer in which a horizontally polarized $H$ photon travels two paths $L$ (left) and $R$ (right) after the first 50:50 beam splitter $BS_1$. This is the `pre-selected' state 
\beq
|\Psi\rangle = \frac{1}{\sqrt{2}}\left(i|L\rangle + |R\rangle\right)|H\rangle
\eeq
The photons along the two arms are recombined on a second 50:50 beam splitter after which two ideal detectors $D_1$ and $D_2$ are placed along the direction of the $R$ and $L$ arms respectively. If the two arms of the interferometer have the same optical length, constructive interference occurs in the direction of the $L$ arm beyond the second beam splitter $BS_2$ and destructive interference in the direction of the $R$ arm beyond the second beam splitter. Hence the detector $D_2$ receives all the photons (the bright port) and $D_1$ none (dark port). This is the standard Mach-Zehnder set up. 

Now, by inserting in the path $R$ a half-wave-plate $HWP$ which flips the polarization from $H$ to $V$ and a phase shifter $P$S which adds a phase $\pi/2$, one obtains the `post-selected' state
\beq
|\Phi\rangle = \frac{1}{\sqrt{2}}\left(|L\rangle|H\rangle + |R\rangle|V\rangle\right).
\eeq
This changes the standard Mach-Zehnder set up and a few photons now enter detector $D_1$.

The final measurement is done using a suitable array of optical elements placed after the second beam splitter. The idea is to get the answer ``yes'' with certainty whenever the system is in the state $|\Phi\rangle$ and the answer ``no'' also with certainty whenever the system is in a state orthogonal to $|\Phi\rangle$. Only those cases in which the answer is ``yes'' are considered. It can be shown using {\em weak measurements} which cause very little disturbance to the system and `nondemolition measurements' (those that do not absorb and/or destroy the photons) on the $L$ and $R$ arms that the photon always travels along the left arm but its polarization always travels along the right arm. Hence, the cat is in the left arm but its grin is in the right arm. Thus, `physical properties can be disembodied from the objects they belong to in a pre- and post-selected experiment'. This conclusion led to many controversies and interpretations but recently a careful experiment has established this disembodiment of position and polarization of an ensemble of single photons in the two arms of an interferometer \cite{home}.

Can many-valued contextual logic resolve this paradox and provide a consistent logical framework for this experiment? The answer is yes, and it lies in the pre- and post-selected states which are both pure states but they are {\em not orthogonal} to each other. This non-orthogonality is basic to weak measurements which give weak values 
\beq
\langle A \rangle_w = \frac{\langle \Phi|A|\Psi\rangle}{\langle\Phi|\Psi\rangle}
\eeq
for any observable $A$. Further, as we have seen, {\em the pre- and post-selected states are prepared under mutually incompatible experimental conditions}: only the detector $D_2$ is bright when the state is $|\Psi\rangle$ but the detector $D_1$ also registers a few photons when the state is $|\Phi\rangle$. The weak and nondemolition measurements of position and polarization are done in the region of the interferometer between the first beam splitter and the half-wave plate HWP (where the state is the pre-selected state $|\Psi\rangle$) and these are considered only when the detector $D_1$ clicks confirming post-selection. Hence, weak measurements are made on a pre- and post-selected ensemble to obtain `weak values' of the position and polarization operators, showing a pure quantum state which is a particle in the $L$ arm and not a particle but a polarization in the $R$ arm. Clearly, it corresponds to the seventh predication ($c_1 T$ and $c_2 F$ and $c_3 U$). Thus, disembodiment of a property of an object in the sense explained above falls within the scope of seven-fold contextual logic. 

That QCC involves contextuality has been noted earlier \cite{hance} but the logical aspects remained unclear.

\section{Psychology of Perception}

While Kothari \cite{kothari} was the first to point out the relevance of {\em sy\={a}dv\={a}da} to quantum mechanics, Mahalanobis \cite{maha} and Haldane \cite{haldane} were the first to point out its relevance in scientific research in general. Haldane wrote, `The search for truth by the scientific method does not lead to complete certainty. Still less does it lead to complete uncertainty. Hence any logical system which allows of conclusions intermediate between certainty and uncertainty should interest scientists. The earliest such system known to me is the Syadvada system of the Jaina philosopher Bhadrabahu (?433-357 B.C.)'. He went on to write,  `I now pass to an example where the saptabhanghinaya is actually applied in scientific research, and which I suspect is not far from what was in Bhadrabahu's mind. In the study of the physiology of the sense organs it is important to determine a threshold. For example a light cannot be seen below a certain intensity, or a solution of a substance which is tasted as bitter when concentrated cannot be distinguished from water when it is diluted. Some experimenters order their subjects to answer ``yes'' or ``no'' to the question ``Is this illuminated?'' or ``Is this bitter?'' If the experimenter is interested in the psychology of perception he will permit the subject also to answer ``It is uncertain'', or some equivalent phrase. The objection to this is that some subjects may do so over a wide range of intensities.' He then shows how, in spite of this objection, seven possibilities ({\em saptabhanginaya}) are exhaustive. He continues,

`I have dealt with a case which arises when the question asked is as simple as possible. Human judgements are generally more complicated. We may attend to the data of several different senses, and of our memories. Thus we arrive at one conclusion from one set of data, and another from another set. We say that wood is hard when compared with clay, soft when compared with iron, indeterminate when compared with similar wood.'

This emphasizes the intrinsic role of contextuality in the psychology of perceptions. With indeterminate answers to questions due to a variety of reasons, probability inevitably enters into experimental psychology. Haldane writes,  

`The close analysis of vision with a dark adapted eye shows that in this case at least Mahalanobis was correct in regarding the saptabhanginaya as foreshadowing modern statistical theory. It appears that when dark adaptation is complete, about five quanta of radiation must arrive within a short time in a small area of the retina before light is reported. Whether they will do so with a given intensity of illuminationn can only be stated as a probability. It is probable, though not by any means certain, that more complicated judgements depend on similar probabilities of events within the central nervous system.' 

\section{{Contextuality and Cognitive Psychology}}

Recent research in cognitive psychology and decision science \cite{bruza1, bruza2, ghose}, has increasingly focused on contextuality--the idea that a cognitive response depends not just on what is measured but also on the context in which it is measured. This challenges traditional cognitive realism, which assumes mental states have predefined, context-independent values, much like the classical notion of physical realism Einstein attempted to retain in quantum mechanics.

Two forms of contextuality are distinguished:

(i) Context-sensitivity (causal), due to signaling or interactions between cognitive variables.

(ii) Genuine contextuality (a-causal), identified through violations of inequalities like Bell or CHSH implied by non-contextuality \cite{khren}.

However, it is difficult to experimentally distinguish between these types, as brain dynamics often involve uncontrollable signaling effects. Moreover, violations of such inequalities are open to multiple interpretations, and may not conclusively indicate contextuality.

In quantum cognition \cite{pot}, cognitive states are modeled using tools from quantum probability (e.g., PVMs, POVMs, quantum instruments), though there is debate about which formalism best captures phenomena like 'order effects' and 'response replicability'. Importantly, quantum cognition remains agnostic toward physical quantum theories of the brain.

A notable recent approach, Contextuality-by-Default (CbD), offers a rigorous framework for identifying contextuality \cite{dz1, dz2, dz3}. It treats the identity of a random variable as dependent on both content and context and emphasizes that variables across different contexts lack joint distributions. Contextuality, in this view, is quantified as the difference between context-free and context-embedded variability in responses.

\section{Concluding Remarks}
Bohr—and Gardner's interpretation of Bohr—place central emphasis on {\em epistemic contextuality}: the meaning of physical quantities in quantum mechanics can only be specified relative to particular experimental setups. They avoid contradictions by treating mutually exclusive descriptions (such as wave and particle behaviour) as complementary, each valid only in its appropriate context.

Reichenbach, followed by Estrada-Gonz\'alez and Cano--Jorge, take a different route by introducing non-classical logic, specifically three-valued logic, to handle quantum indeterminacy. These frameworks account for context-dependence by extending the classical notion of truth, while still preserving logical consistency and avoiding contradictions.

Becker Arenhart and Krause also reject the idea that quantum mechanics involves real contradictions. Instead, they interpret quantum superpositions in terms of potentialities rather than actual, conflicting properties. Their approach accepts contextual constraints on observation, but does not formalize them within a dedicated logical system.

By contrast, da Costa and de Ronde offer a distinctive position: they accept real contradictions as a feature of quantum systems and use paraconsistent logic to handle them. They do not rely on contextuality to resolve paradoxes—instead, contradictions are treated as ontologically real and formally manageable. This marks a clear departure from Bohr's contextualism.

These perspectives can be placed on a spectrum:

Bohr → Reichenbach → Gardner → Estrada-Gonz\'alez and Cano-Jorge: Contextuality is central, but not always formalized; contradictions are avoided by reinterpreting logic, meaning, or semantics.

Becker Arenhart and Krause: Maintain logical consistency; deny the existence of contradictions; accept some contextual constraints but without a formal logical system.

da Costa and de Ronde: Accept contradictions as real; reject the requirement for consistency; do not use contextuality to resolve paradoxes.

Although da Costa and de Ronde reject classical hidden variable theories, their model can be seen as a form of non-classical hidden variable ontology. It posits that contradictory properties exist in quantum systems prior to measurement—though never observed together, these properties are treated as real and logically coherent within a paraconsistent framework. In this sense, their view diverges both from Bohr's epistemic contextualism and from traditional hidden-variable models, by denying the need for determinacy, consistency, or context-sensitive constraints.

This comparative analysis highlights that {\em contextuality} and {\em paraconsistency} constitute fundamentally distinct strategies for addressing the conceptual challenges posed by quantum mechanics, each with profound philosophical implications for our understanding of reality, logic, and knowledge. 

Our seven-valued logic adopts the contextual path and builds on Reichenbach's epistemic logic but brings in the explicit contextual formalism that was missing in all previous frameworks. It also shares with da Costa's approach a non-classical stance on consistency, though in a more formal and philosophically grounded way. It arguably brings Bohr's vision closer to realization than any of the prior attempts by:

(i) Capturing contextuality formally, and

(ii) Providing many-valued distinctions sensitive to experimental setups.

Importantly, in the spirit of Bohr, the scope of this framework extends beyond quantum theory. It offers promising insights in fields such as perceptual psychology and cognitive science, where context-dependence and apparent contradictions are likewise central.


\begin{thebibliography}{0}
\bibitem{bohr}
N. Bohr, ‘Natural Philosophy and Human Cultures’, {\em Nature} {\bf 143}, 268-272 (1939);
reprinted in 'Atomic Physics and Human Knowledge', New York, John Wiley, 23-31 (1958).
\bibitem{ks}
S. Kochen and E. P. Specker, `The Problem of Hidden Variables in Quantum Mechanics', {\em J. of Math and Mechanics} {\bf 17}, 59-87 (1967).
\bibitem{bir}
G. Birkhoff and J. von Neumann, `The Logic of Quantum Mechanics', {\em Annals of Mathematics} {\bf 37}, 823-843 (1936).  
\bibitem{reich}
H. Reichenbach, {\em Philosophic foundations of quantum mechanics}, University of California Press (1944). 
\bibitem{gard}
M. R. Gardner,`Two deviant logics for quantum theory: Bohr and Reichenbach', {\em The British Journal for the Philosophy of Science} {\bf 23} (2), 89-109 (1972). 
\bibitem{costa}
N. C. A. da Costa \& C. de Ronde, `The Paraconsistent Approach to Quantum Superpositions Reloaded:
Formalizing Contradictiory Powers in the Potential Realm', arXiv: 1507.02706 (2015).
\bibitem{ronde}
C. de Ronde, `A Defense of the Paraconsistent Approach to Quantum Superpositions (Answer to Arenhart and Krause)', arXiv: 1401.5186 (2015).
\bibitem{bec}
J. R. Becker Arenhart and D. Krause, `Potentiality and contradiction in quantum mechanics', in A. Koslow \& A. Buchsbaum (Eds.), 'The road to universal logic' (pp. 137-149). {\em Studies in Universal Logic}. Birkhäuser (2015).
\bibitem{est}
L. Estrada-Gonz\'alez and F. Cano-Jorge,  `Revisiting Reichenbach's logic', {\em Synthese}, {\bf 199}, 11821-11845 (2021). 1
\bibitem{ghose}
P. Ghose and S. Patra, {\em An Interdisciplinary Approach to Cognitive Modelling: A Framework Based on Philosophy and Modern Science
},  Routledge; 1st edition (5 December 2023).
\bibitem{graham}
G. Priest, `Coda: Jaina Logic' in {\em The Fifth Corner of Four: An Essay on Buddhist Metaphysics and the Catuskoti}, Oxford University Press (2022).
\bibitem{bel}
N. D. Belnap, `A Useful Four-Valued Logic'. In J.M. Dunn \& G. Epstein (Eds.), {\em Modern Uses of Multiple-Valued Logic}, (pp. 8-37). Dordrecht: Reidel (1977).
\bibitem{schr}
E. Schr\"{o}dinger, `The Present Situation in Quantum Mechanics': A Translation of Schr\"{o}dinger's `Cat Paradox' Paper by J. D. Trimmer, {\em Proceedings of the American Philosophical Society} {\bf 124} (5), 323-338 (1935).
\bibitem{wig}
E. P. Wigner,  `Remarks on the Mind-Body Question'. In Good, I. J. (ed.). {\em The Scientist Speculates: An Anthology of Partly-Baked Ideas}, London: Heinemann. OCLC 476959404. Reprinted In Mehra, Jagdish (ed.). {\em Philosophical Reflections and Syntheses. The Collected Works of Eugene Paul Wigner}. Vol. B/6. Berlin, Heidelberg: Springer. pp. 247-260 (1995).
\bibitem{west}
N. Weststeijn, `Wigner's Friend and Relational Quantum Mechanics: A Reply to Laudisa', {\em Foundations of Physics} {\bf 51}: 86 (2021).
\bibitem{epr}
A. Einstein, B. Podolsky and N. Nathan, `Can Quantum-Mechanical Description of Physical Reality Be Considered Complete?', {\em Phys. Rev.} {\bf 47} (10), 777-780 (1935). 
\bibitem{eins}
A. Einstein, `Autobiographical Notes' in {\em Albert Einstein: Philosopher-Scientist}, The Library of Living Philosophers, ed P. A. Schilpp, Open Court Publishing Company, New York 1949. 
\bibitem{bohr2}
N. Bohr, `Can Quantum-Mechanical Description of Physical Reality be Considered Complete?', {\em Phys. Rev.} {\bf 48} (8), 696-702 (1935).
\bibitem{bohr3}
N. Bohr, {\em The Philosophical Writings of Niels Bohr}, vol. 2, 40-41, Ox Bow Press: Woodbridge, UK, (1987), 
\bibitem{khren}
A. Khrennikov, `Contextuality, Complementarity, Signaling, and Bell Tests', {\em Entropy} {\bf 24} (10), 1380 (2022). 
\bibitem{aha}
Y. Aharonov, D. Rohrlich, S. Popescu, P. Skrzypczyk, `Quantum Cheshire Cats', {\em New J. Phys.} {\bf 15}, 113015 (2013).
\bibitem{home}
S. N. Sahoo, S. Chakraborti, S. Kanjilal, S. R. Behera, D. Home, A. Matzkin and U. Sinha, `Unambiguous joint detection of spatially separated properties of a single photon in the two arms of an interferometer',{\em Nature COMMUNICATIONS PHYSICS} | (2023) 6:203 | https://doi.org/10.1038/s42005-023-01317-7 | www.nature.com
\bibitem{hance}
J. R. Hance, Ming Ji and H. F. Hofman, `Contextuality, coherences, and quantum Cheshire cats', {\em New Journal of Physics} {\bf 25}, 113028 (2023) and references therein.
\bibitem{kothari}
D. S. Kothari, `The Complementarity Principle and Eastern Philosophy' in {\em Niels Bohr: A Centenary Volume}, 325-331, eds A. P. French and P. J. Kennedy, Cambridge, Mass: Harvard University (1985).
\bibitem{maha}
P. C. Mahalanobis, `The foundations of statistics', {\em Dialectica} {\bf 8}(2), 95-111 (1954), reprinted in {\em Sankhy\={a}} {\bf 18} Parts 1 and 2, 183-194 (1957).
\bibitem{haldane}
J. B. S. Haldane, `The Syadvada System of Predication', {\em Sankhy\={a}} {\bf 18}(12), 195-200 (1957).
\bibitem{bruza1}
P. D. Bruza, L. Fell, P. Hoyte, S. Dehdashti, A. Obeid, A. Gibson, and C. Moreira, `Contextuality and context-sensitivity in probabilistic models of cognition', {\em Cognitive Psychology} {\bf 140}, 101529 (2023).
\bibitem{bruza2}
P. Bruza, J. Busemeyer, `Contextual Models of Information Fusion', Queensland University of Technology, (2022).
\bibitem{bruza3}
P. D. Bruza and P. Wittek, `Probabilistic programs for investigating contextuality in human information processing' in {\em Quantum Interaction: 11th International Conference, QI 2018}, Nice, France, September, 2018, Revised Selected Papers 11 (pp. 51-62). Springer International Publishing (2019).  
\bibitem{pot}
E. M. Pothos and J. R. Busemeyer, `Quantum Cognition', {\em Annual Review of Psychology} {\bf 73}, 749-778 (2022).
\bibitem{dz1}
E. N. Dzhafarov, {\em Contents, Contexts, and Basics of Contextuality}, arXiv:2103.07954 [math.PR] (2021).
\bibitem{dz2}
E. N. Dzhafarov, `Assumption-Free Derivation of the Bell-Type Criteria of Contextuality/Nonlocality', {\em Entropy} {\bf 23} (11), 1543 (2021). 
\bibitem{dz3}
I. Basieva, V. H Cervantes, E. N. Dzhafarov, A. Khrennikov, `True contextuality beats direct influences in human decision making', {\em Journal of Experimental Psychology: General} {\bf 148} (11), 1925-1937, (2019).
\end{thebibliography}
\end{document}